\documentclass[a4paper,11pt]{article}
\usepackage{pos}
\usepackage{subfigure} 
\usepackage[absolute,overlay]{textpos} 

\newcommand{\bcrit}{\beta_\text{crit}}
\newcommand{\dslash}[1]{\text{$\not \!\! #1$}}

\title{Finite Temperature Transition in Hyper Stealth Dark Matter using M\"{o}bius Domain Wall Fermions}
\ShortTitle{Finite Temperature Transition in HSDM using MDWF}

\author[a]{Venkitesh Ayyar}
\author[a]{Nobuyuki Matsumoto}
\author[b,c]{Aaron S. Meyer}
\author*[b,c]{Sungwoo Park}
\author[]{Lattice Strong Dynamics (LSD) collaboration}

\affiliation[a]{Hariri Institute for Computing and Computational Science and Engineering, Boston University,
 Boston, MA 02215, USA}
\affiliation[b]{Physical and Life Sciences Division, Lawrence Livermore National Laboratory, Livermore, CA 94550, USA}
\affiliation[c]{Nuclear Science Division, Lawrence Berkeley National Laboratory, Berkeley, CA 94720, USA}

\emailAdd{park49@llnl.gov}

\abstract{ The first-order confinement transition of a strongly
  coupled composite dark matter theory can provide a possible source
  of gravitational waves in the early universe. In this work, on
  behalf of the Lattice Strong Dynamics (LSD) Collaboration, we
  present our recent investigation on the finite temperature
  confinement transition of the one-flavor SU(4) dark gauge theory
  named Hyper Stealth Dark Matter (HSDM). The dark matter candidate in
  this theory is a composite bosonic baryon and can have a remarkably
  low mass of a few GeV. We expect the finite temperature transition
  to be first-order over at least in some finite range of fermionic
  masses and to be a potential source of observable gravitational
  radiation. The finite temperature simulation of one-flavor SU(4) is
  done by using M\"{o}bius Domain wall fermions.  The order of the
  transition and its fermionic mass dependence are explored by
  monitoring the Polyakov loop, chiral condensate and topological
  charge using three lattice volumes at $N_t=8$.}

\FullConference{The 41st International Symposium on Lattice Field Theory (LATTICE2024)\\
 28 July - 3 August 2024\\
Liverpool, UK\\}

\begin{document}

\begin{textblock}{20}(14.0,1.70)
LLNL-PROC-872252
\end{textblock}

\maketitle

\section{Introduction}
\label{sec:into}

In recent years, the early universe has become a phenomenological
laboratory to study new physics.  In Ref.~\cite{Fleming:2024flc}, the
one-flavor SU(4) gauge theory was proposed as a model for the dark
matter, which was named as the Hyper Stealth Dark Matter (HSDM).  The
dark matter candidate in this model is the lightest baryon composed of
four dark-quarks, which can be as light as a few GeV.  In the
high-temperature phase of the universe, the HSDM is in a
dark-quark-dark-gluon plasma, and as the universe cools down, the HSDM
undergoes a phase transition into a confined phase of dark-hadrons.
This phase transition, which is triggered by non-perturbative dynamics
of the strongly-interacting theory, gives us another phenomenological
interest as it can generate a stochastic background of gravitational
waves if the transition is first-order \cite{Schwaller:2015tja} (see,
e.g., \cite{Caprini:2019egz,LISACosmologyWorkingGroup:2022jok} for
reviews; a possibility of detecting gravitational waves from crossover
is also discussed \cite{Escriva:2024ivo}).

In this contribution, we study the confinement transition in the
strongly-interacting SU(4) sector of this one-flavor model as a
continuation of our previous report \cite{Ayyar:2024dmt}.  In
particular, we determine the order of the phase transition with
various dark-quark masses: $am\in \{0.01, 0.05, 0.1, 0.2, 0.3, 0.4,
\infty\}$, including the quenched theory, $am=\infty$.  We use the
Wilson gauge action (with the conventional coupling $\beta$) and the
M\"{o}bius Domain Wall Fermion (MDWF)
\cite{Kaplan:1992bt,Brower:2012vk}.  We used three lattice volumes
$N_s^3\times N_t$ with $N_s=$16, 24, and 32 at fixed $N_t=8$ and
$L_s=16$.  Gauge configurations are generated by the HMC with the
exact one flavor algorithm (EOFA) \cite{Chen:2014hyy}.  Calculations
are performed on LLNL clusters using Grid \cite{Boyle:2015tjk}.

As is the case for the one-flavor theory with $N_c=3$
\cite{Alexandrou:1998wv}, we expect the deconfinement transition to be
first-order with heavy dark-quark masses and crossover with light
dark-quark masses, separated by a second-order transition point.  We
study the critical coupling $\beta=\bcrit$ for various dark-quark
masses by using the Polyakov loop as the order parameter.  Note that,
in the one-flavor theory, there is no spontaneous chiral symmetry
breaking as $U_A(1)$ is broken explicitly by the chiral anomaly.
Accordingly, the lightest meson $\eta'$ acquires a mass of the order
of the intrinsic scale of the theory, and it is expected to have no
chiral phase transition \cite{Pisarski:1983ms,Alexandrou:1998wv}.  In
this regard, we also investigate the chiral condensate and the
topological charge around the confinement transition point $\beta_{\rm
  crit}$.

\section{M\"{o}bius Domain Wall Fermion}
\label{sec:mdwf}

As we have mentioned in the Introduction, the relation between chiral
behavior and the deconfinement transition is of theoretical interest.
In this study, we use the M\"{o}bius domain-wall fermion (MDWF)
\cite{Kaplan:1992bt,Brower:2012vk} to systematically control the
chiral symmetry breaking of the theory.  As is well understood (see,
e.g., Ref.~\cite{RBC:2008cmd} and reference therein), the breaking of
the chiral symmetry due to finite $L_s$ for domain-wall fermions can
be quantified by the residual mass $m_\text{res}$, which is related to
the small eigenvalues of the hermitian Dirac operator:
$H_4(M_5)=\gamma_5\dslash{D}_W(-M_5)$, where $M_5$ is the domain-wall
height.  In this section, we briefly describe this residual breaking
of the chiral symmetry in our ensembles.

\begin{figure}[tb]
  \center
  \includegraphics[width=0.32\linewidth]{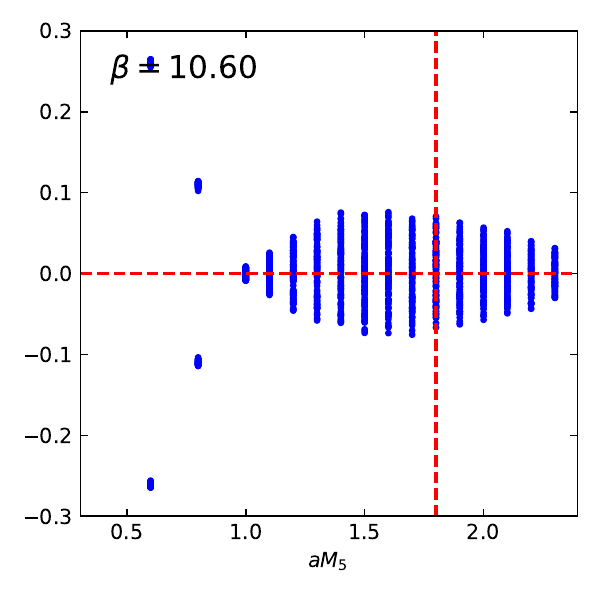}
  \includegraphics[width=0.32\linewidth]{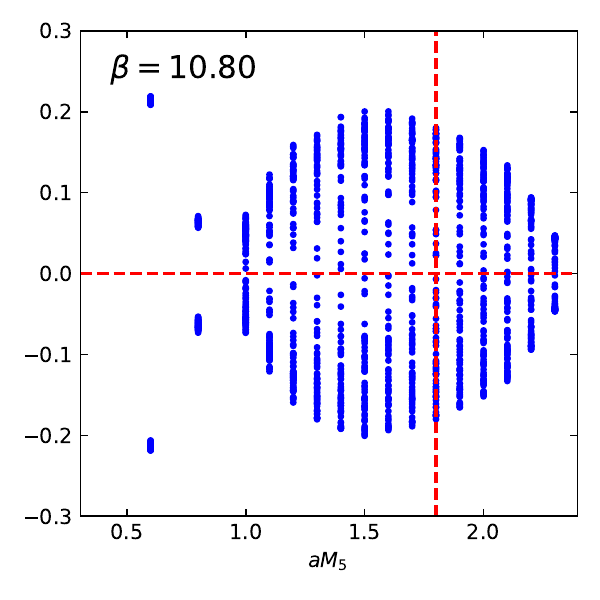}
  \includegraphics[width=0.32\linewidth]{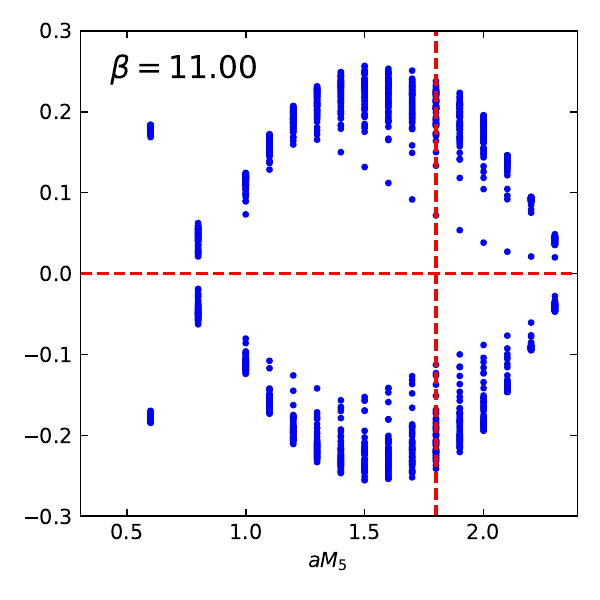}
  \caption{The ten smallest magnitude eigenvalues of the
    four-dimensional hermitian Dirac operator: $H_4(M_5)=\gamma_5
    \dslash{D}_W(-M_5)$, calculated with ten thermalized
    configurations for $\beta=10.6, 10.8, 11.0$ (from left to right)
    with the volume $24^3\times 8$ and the bare quark mass $am=0.1$.
    The red vertical dotted line indicates $aM_5=1.8$, which is the
    parameter used in the HMC for $am=0.1$ and $0.4$.  }
  \label{fig:eye}
\end{figure}

\begin{figure}[t]
  \centering
  \includegraphics[width=0.45\linewidth]{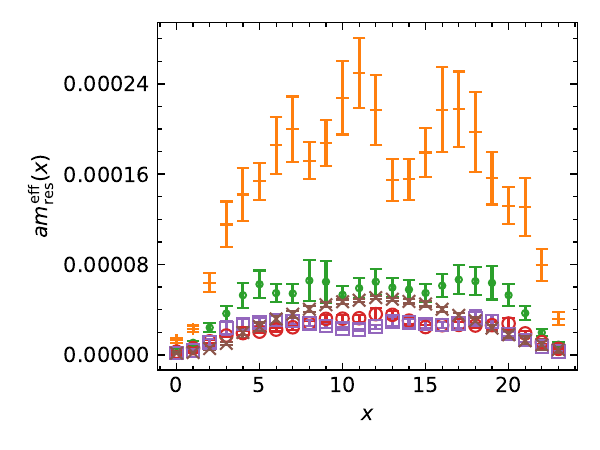}
  \includegraphics[width=0.45\linewidth]{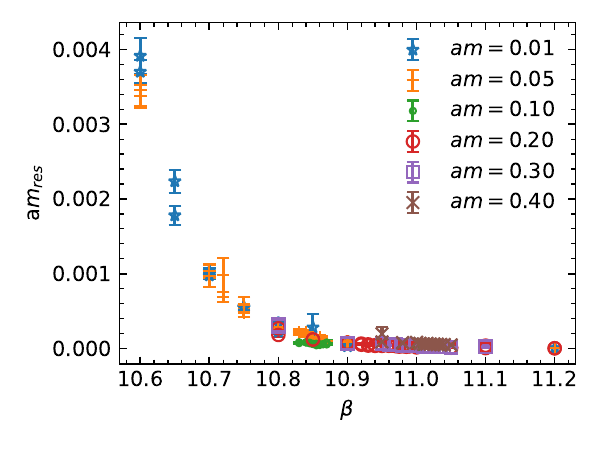}
  \caption{(Left) The effective residual mass $am_{\rm res}^{\rm
      eff}(x) =\langle J_{5q}(x)P(0)\rangle / \langle P(x)P(0)\rangle
    $ as a function of the spatial separation $x$, evaluated at
    $\bcrit$ for each quark mass. (Right) The residual mass $m_{\rm
      res}$ determined from the midpoint of $am_{\rm res}^{\rm
      eff}(x)$, plotted as a function of $\beta$. Both plots are
    calculated with the $24^3\times 8$ ensembles. }
  \label{fig:m_res}
\end{figure}

Figure~\ref{fig:eye} shows the first ten smallest magnitude
eigenvalues of $H_{4}(M_{5})$ as a function of $M_{5}$ with the bare
quark mass $am$ = 0.1.  We show results for the three representative
cases with $am=0.1$: $\beta=10.6$, which corresponds to the deconfined
phase; $\beta=10.8$, around the critical point; and $\beta=11.0$, the
confined phase. The correspondence between $\beta$ and the phases will
be given in Sec.~\ref{sec:PL}.
The red vertical line in the figure marks $M_5=1.8$, which is used in
the HMC to generate the gauge configurations for $am=0.1$ and
$0.4$. For simulations with other quark masses
$am\in\{0.05,0.2,0.3\}$, we set $M_{5}=1.5$ to improve chiral
behavior, based on Fig.~\ref{fig:eye}. To further confirm that the
breaking of the chiral symmetry according to finite $L_s$ is well
controlled, we calculate the residual mass from the axial Ward
identity \cite{CP-PACS:2000fmi}: $m_\text{res} =
\allowbreak\lim_{x\to\infty} \langle J_{5q} (x)P(0)\rangle /
\allowbreak \langle P(x)P(0)\rangle$, where $P$ is the pseudoscalar
meson operator and $J_{5q}$ the five-dimensional flavor non-singlet
axial current.
In the left panel of Fig.~\ref{fig:m_res}, we show the effective
residual mass $m_\text{res}^\text{eff}$ as a function of the spatial
separation $x$ for various quark masses at the critical coupling
$\bcrit$ which will be determined in Sec.~\ref{sec:PL}.
The value of $m_\text{res}$ is given at the midpoint.  The effect of
of the residual mass compared to the bare quark mass,
$m_\text{res}/m$, is sub-percent for the parameters of interest and is
therefore negligible.

\section{Gauge ensemble}

\begin{figure}[htb]
  \center
  \includegraphics[width=0.32\linewidth]{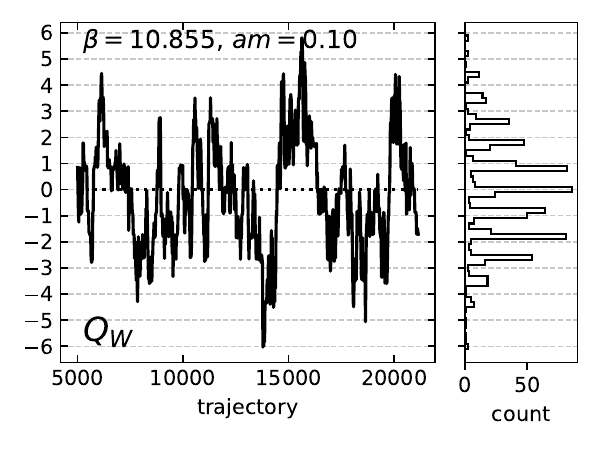}
  \includegraphics[width=0.32\linewidth]{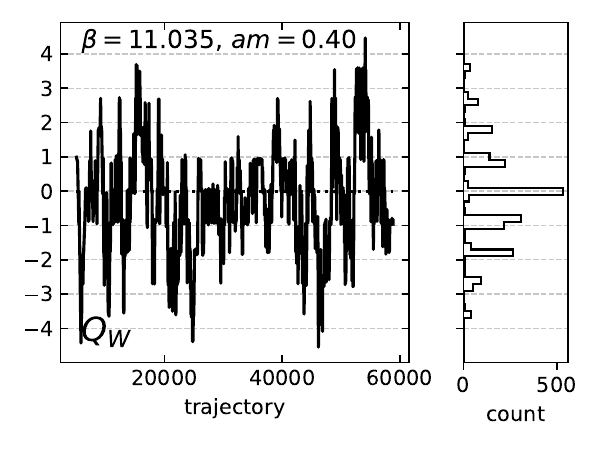}
  \includegraphics[width=0.32\linewidth]{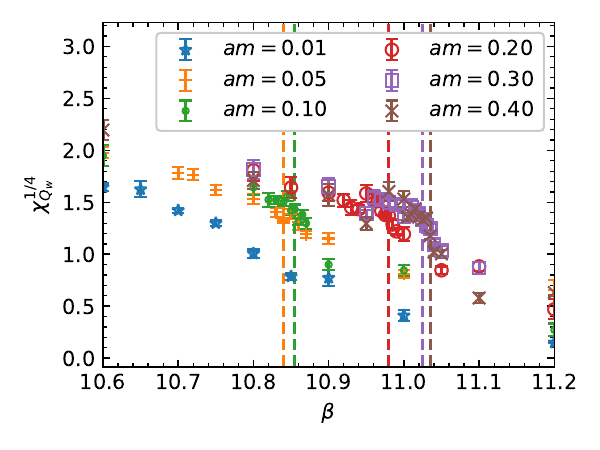}
  \caption{Monte Carlo time history of the Wilson-flowed topological
    charge $Q_W$, and the corresponding histogram calculated at
    $\bcrit$ for the masses: (left) $am=0.1$ and (middle) $am=0.4$.
    In the right panel, we show the topological susceptibility as a
    function of $\beta$ for the six different quark masses: $m=0.01,
    0.05, 0.1, 0.2, 0.3, 0.4$.  Vertical dotted lines are drawn at
    $\beta=\bcrit$ to indicate the transition point for each mass with
    the same color used for the susceptibility.  }
  \label{fig:Q_W}
\end{figure}

To ensure thermalization, we create two streams with a cold start from
the unit gauge field ($U_\mu=1$) and a hot start from a random gauge
field for each $\beta$ and the quark mass $am$.  The sample size
varies from 3000 to $O(10^5)$ (for the quenched case, up to
$O(10^6)$), taking into account the diverging autocorrelation of the
Polyakov loop and the topological charge.  Figure \ref{fig:Q_W} shows
the Monte Carlo time history and the histogram of the Wilson-flowed
topological charge $Q_W$, where we adopt the clover leaf definition
for the field strength.  The Wilson flow time is fixed to $2.0$ with
the step size $0.01$.  Though we see long autocorrelation, the sample
size is large enough to observe a decent number of tunnelings in the
ensemble, and as a result we obtain a symmetric distribution.  With
the current statistics, we do not observe a peak structure for the
chiral susceptibility $\chi_{Q_W}$ (see the right-most panel in
Fig.~\ref{fig:Q_W}) around the critical coupling $\bcrit$.

\section{Deconfinement transition}
\label{sec:PL}

We study the Polyakov loop ($PL$) as an order parameter for the
deconfinement phase transition.  Though the dynamical quark explicitly
breaks the center symmetry, we expect it to still serve as an order
parameter for large quark masses.
\begin{figure}[htb]
  \centering
  \includegraphics[width=0.6\linewidth]{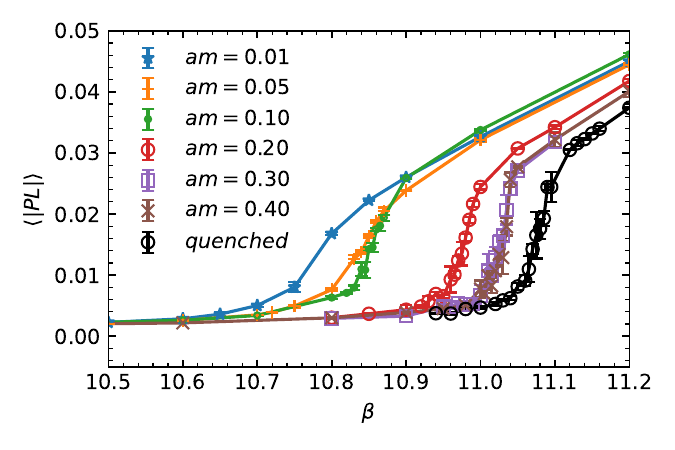}
  \caption{Polyakov Loop absolute value versus $\beta$ on $24^3\times 8$ ensembles.
  }
  \label{fig:pl_abs_24}
\end{figure}
Figure \ref{fig:pl_abs_24} shows the expectation value of the absolute
value, $\langle \vert PL \vert \rangle$, as a function of $\beta$ for
various quark masses.  The vanishing expectation values $\langle \vert
PL \vert \rangle = 0$ at small $\beta$ indicate the confined phase,
while the nonzero values $\langle \vert PL \vert \rangle \neq 0$ at
large $\beta$ signify the deconfined phase.  A diverging slope in the
transition region in the infinite volume limit implies a phase
transition.

\begin{figure}[htb]
  \centering
  \includegraphics[width=0.45\linewidth]{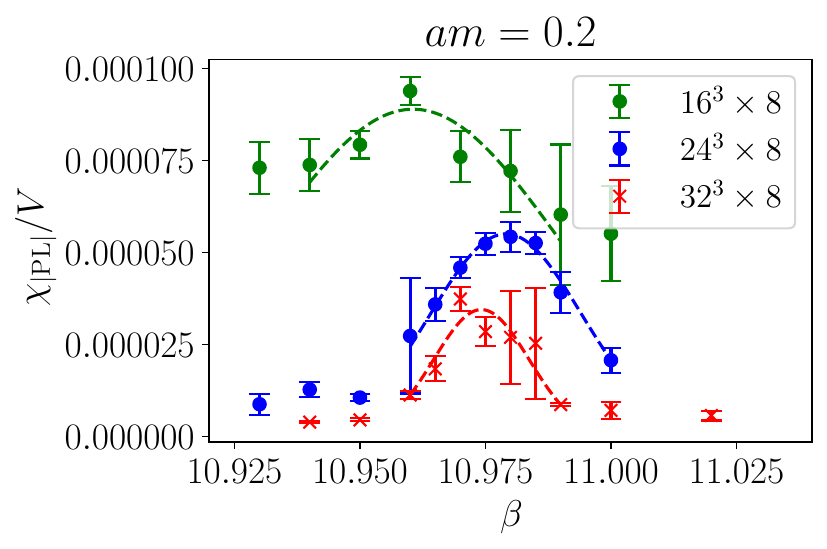}
  \includegraphics[width=0.45\linewidth]{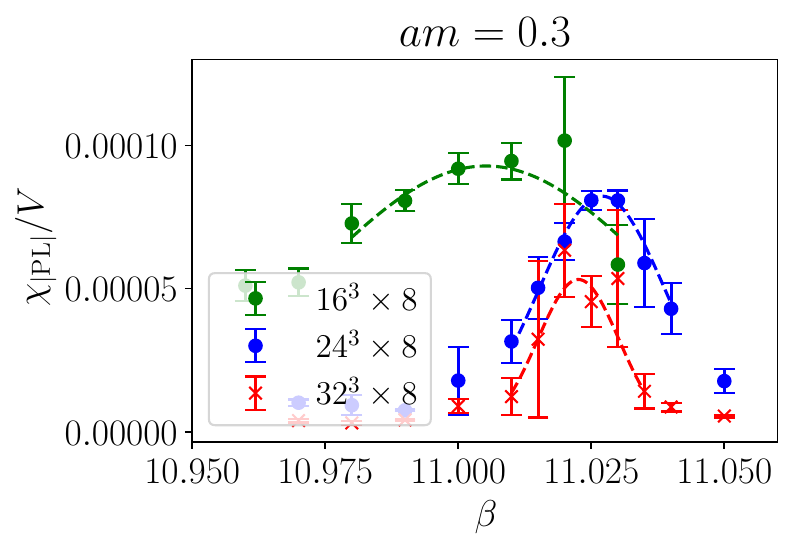}
  \\\includegraphics[width=0.45\linewidth]{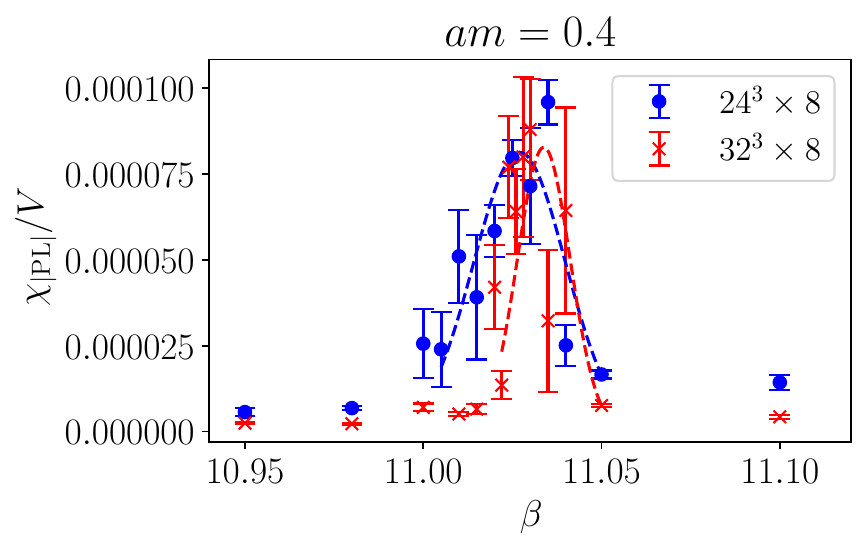}
  \includegraphics[width=0.45\linewidth]{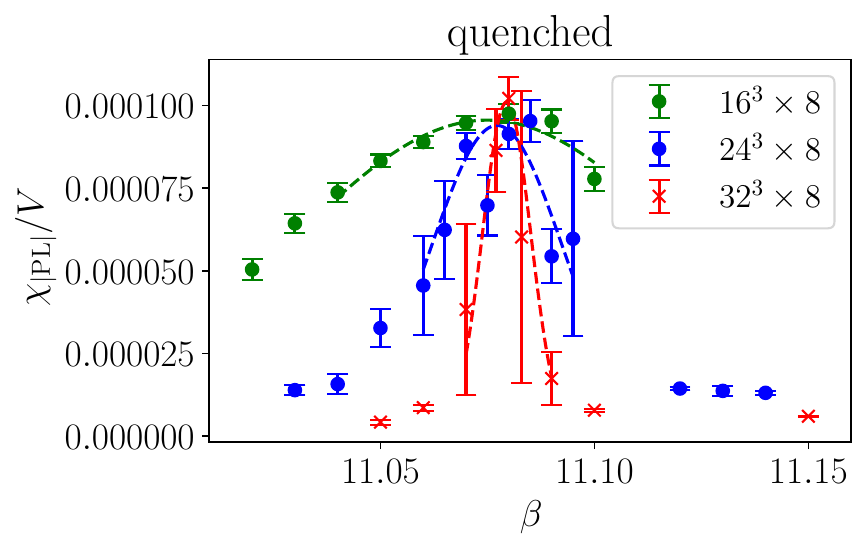}
  \caption{ The Polyakov loop susceptibility normalized by the spatial
    volume, $\chi_{\rm |PL|}/V$, where $V=N_s^3$.  Green, blue, and
    red points show the results for $16^3\times 8$, $24^3\times 8$,
    and $32^3\times 8$, respectively.  The ensembles with hot and cold
    starts are combined in this plot.  As a preliminary study, the
    bell-shaped peaks are fitted in the Gaussian form, whose results
    are drawn with dashed lines.  We see that the peak value of the
    susceptibility scales linearly with volume for $m=0.4$ and the
    quenched case, signifying first-order phase transition.  }
  \label{fig:pl_chi_24}
\end{figure}

To further discuss the phase transition, we plot in
Fig.~\ref{fig:pl_chi_24} the susceptibility of the Polyakov loop,
which shows a peak structure at the transition region.  The order of
the phase transition can then be identified by studying the finite
volume scaling of the height $\chi_{|PL|}^\text{max}$ of the peak, for
which we make the ansatz: $\chi_{|PL|}^\text{max}\propto N_s^{3b}$.
If the phase transition is first-order, the exponent $b$ is expected
to be 1 \cite{Binder:1984llk, Fukugita:1989yb}.  Although the notion
of the critical temperature is obscure when the transition is
crossover, we write as $\beta_{\rm crit}$ the location of the peak in
the susceptibility.  As a preliminary study, we here determine the
peak value $\chi_{|PL|}^\text{max}$ by fitting the bell-shaped peak in
the Gaussian form.  The obtained values for the exponent are
$b=0.46(10)$ for $am=0.2$, $b=0.49(19)$ for $am=0.3$, $b=1.02(19)$ for
$am=0.4$, and $b=1.009(23)$ for the quenched theory.  We do not use
the results of $16^3\times 8$ in estimating $b$ for $am=0.2, 0.3$ as
it is deviates significantly from the scaling ansatz, in relation to
which we can observe in Fig.~\ref{fig:pl_chi_24} that $16^3\times 8$
gives a significantly different value of $\beta_{\rm crit}$.\footnote{
We comment that the $16^3\times 8$ ensembles shown in this figure are
generated with $M_5=1.8$ while the $24^3\times 8$ and $32^3\times 8$
ensembles with $M_5=1.5$.  While the parameters are to be unified, we
expect the resulting difference in the observable to be small as
described in Sec.~\ref{sec:mdwf}.} 
for $16^3\times 8$ with $am=0.4$.  Our analysis implies that the
second-order point lies somewhere around $am=0.3$ and $0.4$.  Most
importantly, we find a finite dark-quark mass that is consistent with
a first-order transition that can source
gravitational waves.

\begin{figure}[htb]
\centering
  \includegraphics[width=0.25\linewidth]{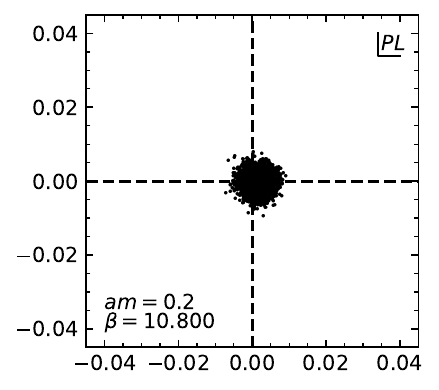}
  \includegraphics[width=0.25\linewidth]{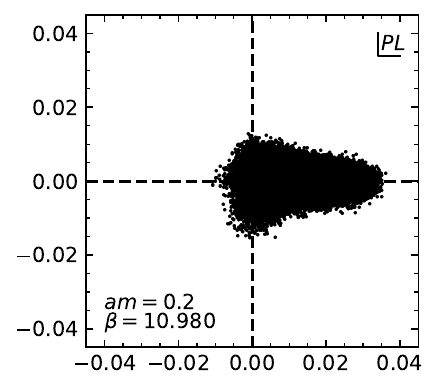}
  \includegraphics[width=0.25\linewidth]{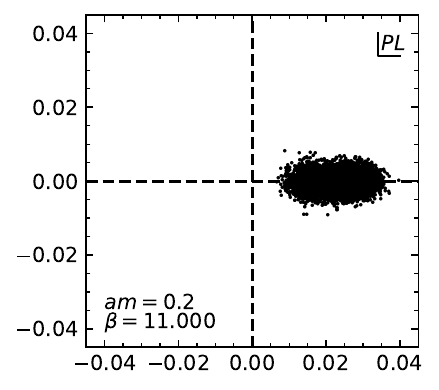}
  \\
  \includegraphics[width=0.25\linewidth]{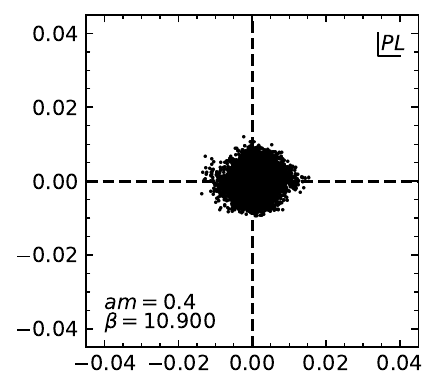}
  \includegraphics[width=0.25\linewidth]{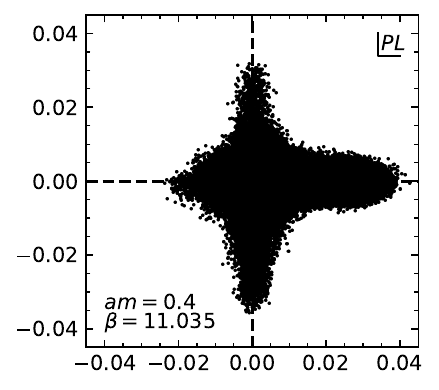}
  \includegraphics[width=0.25\linewidth]{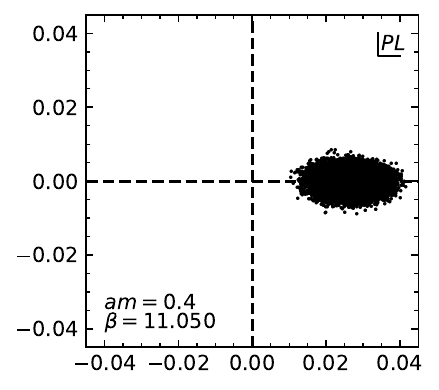}
  \\
  \includegraphics[width=0.25\linewidth]{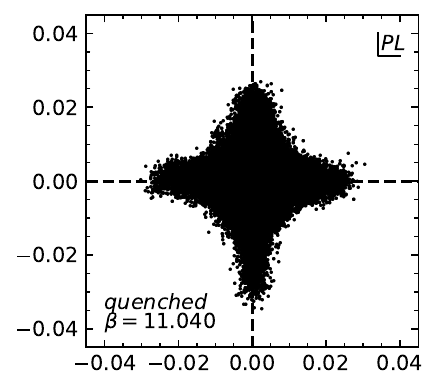}
  \includegraphics[width=0.25\linewidth]{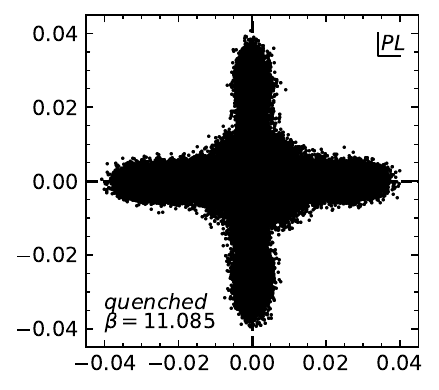}
  \includegraphics[width=0.25\linewidth]{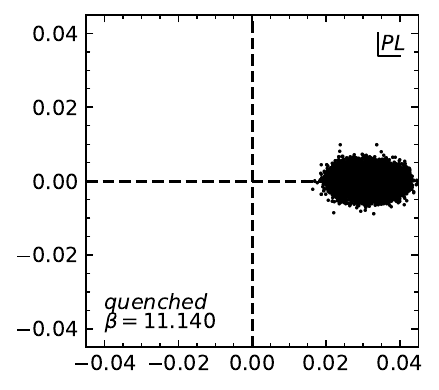}
  \caption{ The scattered plot of the spatially averaged Polyakov loop
    in the complex plane for the masses $am=0.2, 0.4$, and the
    quenched theory (from top to the bottom rows) on the $24^3\times
    8$ lattice.  We choose three $\beta$ values for each $am$: below,
    on top of, and above the coupling $\beta_{\rm crit}$ (from left to
    right).  }
  \label{fig:pl_comp_24}
\end{figure}

It is interesting to see how the distribution of the Polyakov loop
evolves across the phases.  Figure \ref{fig:pl_comp_24} shows the
scattered plot of the Polyakov loop in the complex plane before, in
the middle of, and after the transition region, from left to right.
In the quenched case (bottom row), where the $Z_4$ symmetry is
preserved classically, we observe that the distribution of the
Polyakov loop exhibits the $Z_4$ symmetry in the confined phase
(bottom left) and it is broken spontaneously in the deconfinement
phase (bottom right).  Correspondingly, we observe $\langle PL \rangle
= 0$ in the confined phase and $\langle PL \rangle \neq 0$ in the
deconfined phase.
With the dynamical dark-quarks (top and middle rows), where the $Z_4$
symmetry is explicitly broken, we see that the distribution in the
confined phase is localized around zero (top/middle left), resulting
in the expectation value $\langle PL\rangle = 0$.  As we increase
$\beta$, we observe a departure from the origin, moving towards the
positive real axis (top/middle right), which we interpret as that the
system has transited into the deconfined phase, giving $\langle PL
\rangle \neq 0$.

\begin{figure}[htb]
  \centering
    \includegraphics[width=0.48\linewidth]{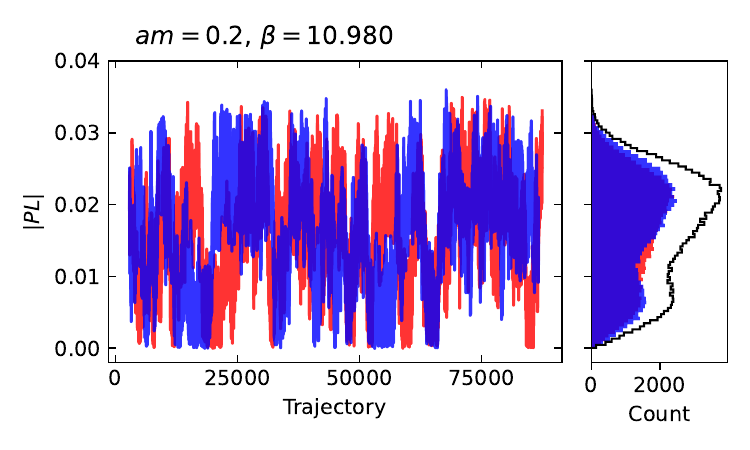}
    \includegraphics[width=0.48\linewidth]{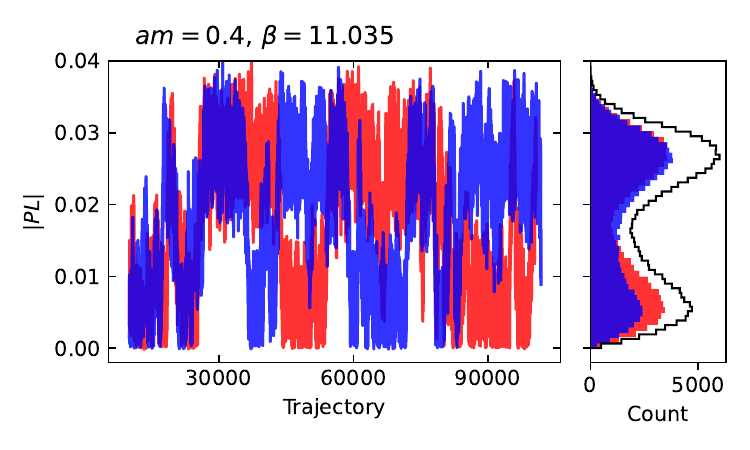}
\caption{The Monte Carlo time history of the absolute value of the
  Polyakov loop, $|PL|$, and its histogram at $\bcrit$ for the masses
  below and above the second-order transition point: $am=0.2, 0.4$ on
  the $24^3\times 8$ lattice. The red and blue colors represent the
  hot and cold start streams, respectively, while the black curve
  shows the histogram that combines data from both streams.  }
  \label{fig:pl_hist}
\end{figure}

To further scrutinize the transition, we look into the histogram.  In
Fig.~\ref{fig:pl_hist}, we plot the Monte Carlo time history and the
histogram of $|PL|$ at $\beta_{\rm crit}$ for $m=0.2$ (in the
crossover region) and $m=0.4$ (in the first-order region).  Red and
blue colors represent the hot and cold start streams, respectively.
The first-order nature of the phase transition at $am=0.4$ can be seen
as the double-peaked structure in the histogram.  We can further
confirm the tendency that the separation between the two peaks becomes
obscure as we decrease the dark-quark mass.

\section{Chiral susceptibility}

\begin{figure}[htb]
  \centering
  \includegraphics[width=0.32\linewidth]{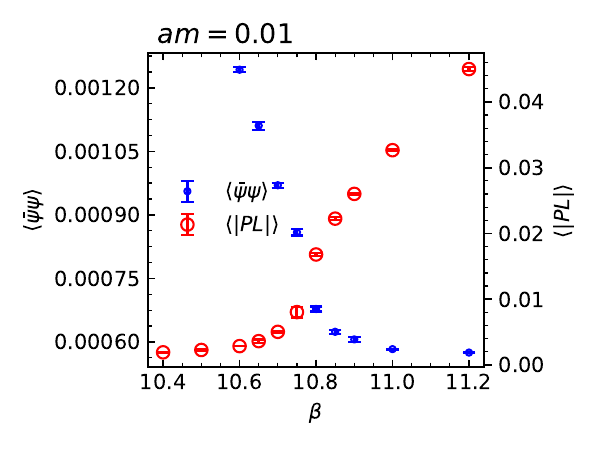}
  \includegraphics[width=0.32\linewidth]{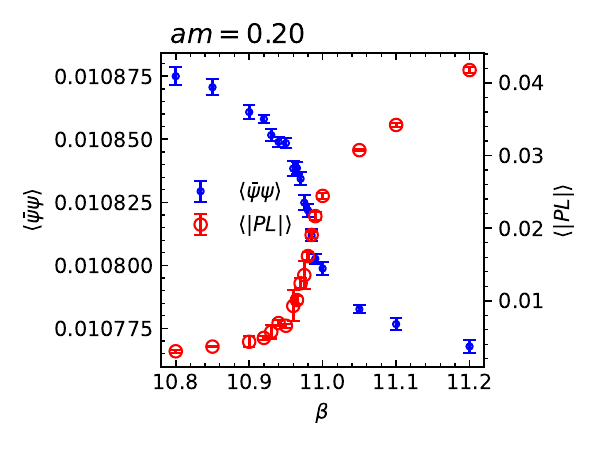}
  \includegraphics[width=0.32\linewidth]{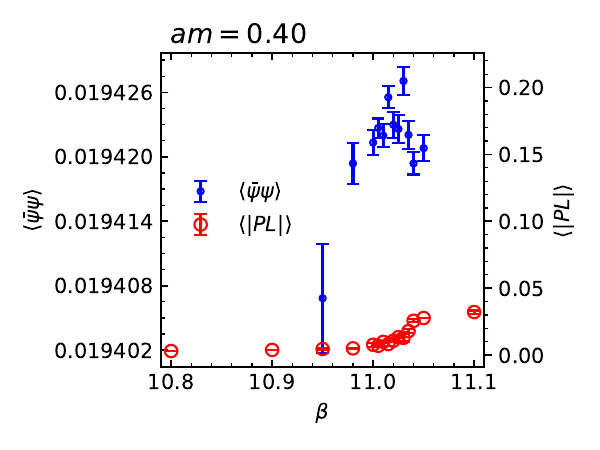}  
  \includegraphics[width=0.32\linewidth]{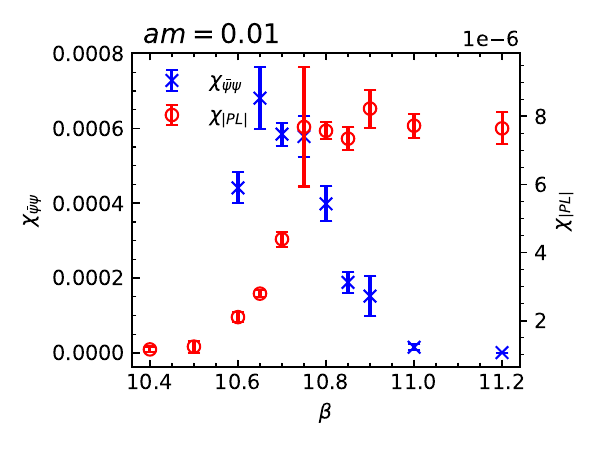}
  \includegraphics[width=0.32\linewidth]{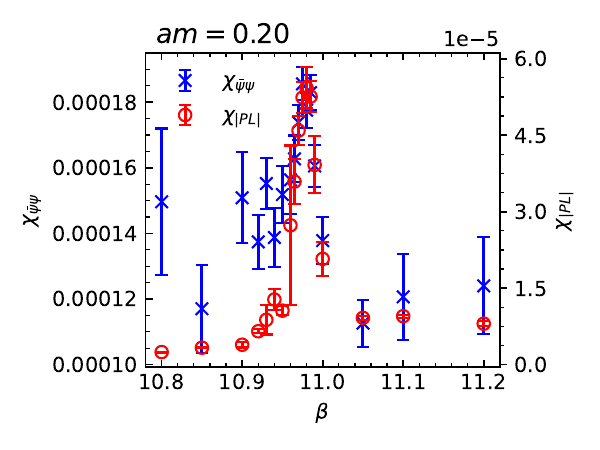}
  \includegraphics[width=0.32\linewidth]{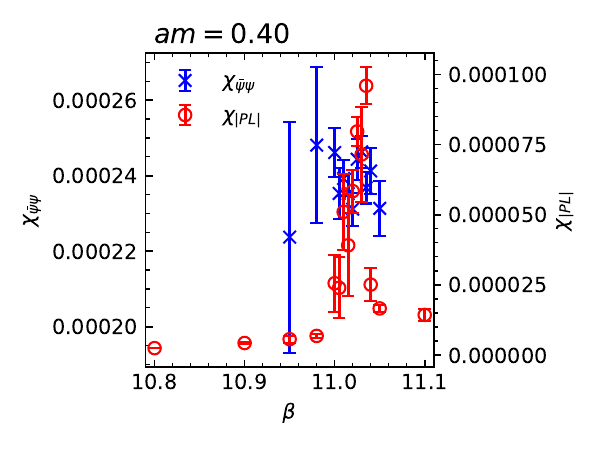}
  \caption{(Top) The chiral condensate and the Polyakov loop as a
    function of $\beta$. (Bottom) The disconnected chiral
    susceptibility and the Polyakov loop susceptibility as a function
    of $\beta$.  The lattice size is fixed to $24^3\times 8$ in this
    figure, while the dark-quark mass is varied as $am=0.01, 0.2, 0.4$
    from left to right.}
  \label{fig:pl_pbp_24}
\end{figure}

As mentioned in the Introduction, the chiral symmetry is broken in the
one-flavor theory by the axial anomaly.  Consequently, unlike in QCD,
chiral symmetry may not be restored in the large $\beta$ limit.  By
using the chiral condensate $\langle \bar\psi \psi\rangle$ as the
order parameter for the chiral phase transition, we can check this
numerically in our model.
The top three panels in Fig.~\ref{fig:pl_pbp_24} show the behavior of
$\langle \bar\psi \psi\rangle$ around $\beta_{\rm crit}$ for the
masses $am=0.01, 0.2, 0.4$, drawn together with $\langle | PL |
\rangle$ for comparison.  We use a noisy estimator to calculate the
condensate.  It is interesting to observe a steep slope around
$\beta_{\rm crit}$ even for $\langle \bar\psi \psi\rangle$.  In the
bottom three panels, we display the disconnected chiral
susceptibility, $\chi_{\bar\psi\psi} = N_s^3N_t (\langle (\bar\psi
\psi)^2\rangle_\text{disc} - \langle \bar\psi \psi\rangle^2)$,
together with $\chi_{|PL|}$, in which the coincidental peak locations
can be confirmed (especially for $am=0.2$).
Further study on their relative locations as well as the finite volume
scaling for the chiral susceptibility are interesting in order to
understand the non-perturbative dynamics of the theory.  The study is
in progress with improved statistics.

\section{Conclusion and outlook}

In this contribution, we gave an update from Ref.~\cite{Ayyar:2024dmt}
on a strongly-interacting composite dark matter theory, the HSDM
\cite{Fleming:2024flc}; an SU(4) gauge theory with one flavor of
M\"{o}bius domain-wall fermion.  We explored the thermodynamics of
this theory at various quark masses and $\beta$ values by computing
the topological charge, the Polyakov loop and the chiral condensate.
We found that the transition is first-order at a large but moderate
dynamical fermion mass.  More detailed analysis with improved
statistics is in progress, as well as zero temperature ensemble
generation at $\bcrit$ for scale-setting in the theory by measuring
the SU(4) meson and baryon masses.

\begin{acknowledgments}

This work was performed under the auspices of the U.S. Department of
Energy by Lawrence Livermore National Laboratory under Contract
DE-AC52-07NA27344. ASM is supported in part by Neutrino Theory Network
Program Grant DE-AC02-07CHI11359, and U.S. Department of Energy Award
DE-SC0020250. SP acknowledges the support from the ASC COSMON
project. SP, ASM and P.~M.~Vranas would like to thank Scott Futral of
LLNL for his support and early access to LLNL's exascale systems,
Tuolumne and El Capitan where most of the numerical simulations were
performed.  NM is supported in part by the Scientific Discovery
through Advanced Computing (SciDAC) program, ``Multiscale
acceleration: Powering future discoveries in High Energy Physics''
under FOA LAB-2580 funded by DOE, Office of Science, and DOE under
Award DE-SC0015845.  Results presented in this contribution were
produced using Grid \cite{Boyle:2015tjk}.

\end{acknowledgments}

\bibliographystyle{JHEP}
\bibliography{ref} 

\providecommand{\href}[2]{#2}\begingroup\raggedright\begin{thebibliography}{10}

\bibitem{Fleming:2024flc}
G.~T. Fleming, G.~D. Kribs, E.~T. Neil, D.~Schaich, and P.~M. Vranas, {\it
  {Hyper Stealth Dark Matter and Long-Lived Particles}},
  \href{http://xxx.lanl.gov/abs/2412.14540}{{\tt 2412.14540}}.

\bibitem{Schwaller:2015tja}
P.~Schwaller, {\it {Gravitational Waves from a Dark Phase Transition}},  {\em
  Phys. Rev. Lett.} {\bf 115} (2015), no.~18 181101,
  [\href{http://xxx.lanl.gov/abs/1504.07263}{{\tt 1504.07263}}].

\bibitem{Caprini:2019egz}
C.~Caprini {\em et~al.}, {\it {Detecting gravitational waves from cosmological
  phase transitions with LISA: an update}},  {\em JCAP} {\bf 03} (2020) 024,
  [\href{http://xxx.lanl.gov/abs/1910.13125}{{\tt 1910.13125}}].

\bibitem{LISACosmologyWorkingGroup:2022jok}
{\bf LISA Cosmology Working Group} Collaboration, P.~Auclair {\em et~al.}, {\it
  {Cosmology with the Laser Interferometer Space Antenna}},  {\em Living Rev.
  Rel.} {\bf 26} (2023), no.~1 5,
  [\href{http://xxx.lanl.gov/abs/2204.05434}{{\tt 2204.05434}}].

\bibitem{Escriva:2024ivo}
A.~Escriv\`a, R.~Inui, Y.~Tada, and C.-M. Yoo, {\it {LISA forecast on a smooth
  crossover beyond the standard model through the scalar-induced gravitational
  waves}},  {\em Phys. Rev. D} {\bf 111} (2025), no.~2 023528,
  [\href{http://xxx.lanl.gov/abs/2404.12591}{{\tt 2404.12591}}].

\bibitem{Ayyar:2024dmt}
{\bf LSD} Collaboration, V.~Ayyar, {\it {Exploring Composite Dark Matter with
  an SU(4) gauge theory with 1 fermion flavor}},  {\em PoS} {\bf LATTICE2023}
  (2024) 102, [\href{http://xxx.lanl.gov/abs/2402.07362}{{\tt 2402.07362}}].

\bibitem{Kaplan:1992bt}
D.~B. Kaplan, {\it {A Method for simulating chiral fermions on the lattice}},
  {\em Phys. Lett. B} {\bf 288} (1992) 342--347,
  [\href{http://xxx.lanl.gov/abs/hep-lat/9206013}{{\tt hep-lat/9206013}}].

\bibitem{Brower:2012vk}
R.~C. Brower, H.~Neff, and K.~Orginos, {\it {The M\"obius domain wall fermion
  algorithm}},  {\em Comput. Phys. Commun.} {\bf 220} (2017) 1--19,
  [\href{http://xxx.lanl.gov/abs/1206.5214}{{\tt 1206.5214}}].

\bibitem{Chen:2014hyy}
{\bf TWQCD} Collaboration, Y.-C. Chen and T.-W. Chiu, {\it {Exact Pseudofermion
  Action for Monte Carlo Simulation of Domain-Wall Fermion}},  {\em Phys. Lett.
  B} {\bf 738} (2014) 55--60, [\href{http://xxx.lanl.gov/abs/1403.1683}{{\tt
  1403.1683}}].

\bibitem{Boyle:2015tjk}
P.~Boyle, A.~Yamaguchi, G.~Cossu, and A.~Portelli, {\it {Grid: A next
  generation data parallel C++ QCD library}},
  \href{http://xxx.lanl.gov/abs/1512.03487}{{\tt 1512.03487}}.

\bibitem{Alexandrou:1998wv}
C.~Alexandrou, A.~Borici, A.~Feo, P.~de~Forcrand, A.~Galli, F.~Jegerlehner, and
  T.~Takaishi, {\it {The Deconfinement phase transition in one flavor QCD}},
  {\em Phys. Rev. D} {\bf 60} (1999) 034504,
  [\href{http://xxx.lanl.gov/abs/hep-lat/9811028}{{\tt hep-lat/9811028}}].

\bibitem{Pisarski:1983ms}
R.~D. Pisarski and F.~Wilczek, {\it {Remarks on the Chiral Phase Transition in
  Chromodynamics}},  {\em Phys. Rev. D} {\bf 29} (1984) 338--341.

\bibitem{RBC:2008cmd}
{\bf RBC, UKQCD} Collaboration, D.~J. Antonio {\em et~al.}, {\it {Localization
  and chiral symmetry in three flavor domain wall QCD}},  {\em Phys. Rev. D}
  {\bf 77} (2008) 014509, [\href{http://xxx.lanl.gov/abs/0705.2340}{{\tt
  0705.2340}}].

\bibitem{CP-PACS:2000fmi}
{\bf CP-PACS} Collaboration, A.~A. Khan {\em et~al.}, {\it {Chiral properties
  of domain wall quarks in quenched QCD}},  {\em Phys. Rev. D} {\bf 63} (2001)
  114504, [\href{http://xxx.lanl.gov/abs/hep-lat/0007014}{{\tt
  hep-lat/0007014}}].

\bibitem{Binder:1984llk}
K.~Binder and D.~P. Landau, {\it {Finite-size scaling at first-order phase
  transitions}},  {\em Phys. Rev. B} {\bf 30} (1984), no.~3 1477.

\bibitem{Fukugita:1989yb}
M.~Fukugita, M.~Okawa, and A.~Ukawa, {\it {Order of the Deconfining Phase
  Transition in SU(3) Lattice Gauge Theory}},  {\em Phys. Rev. Lett.} {\bf 63}
  (1989) 1768.

\end{thebibliography}\endgroup

\end{document}